\documentstyle[aps,prl,multicol]{revtex}
\tighten
\draft
\begin{document}
\title{Parity Problem With A Cellular Automaton Solution} 
\author{K. M. Lee, Hao Xu and H. F. Chau\footnote{Corresponding author,
 Email address: hfchau@hkusua.hku.hk}}
\address{Department of Physics, University of Hong Kong, Pokfulam Road, Hong
 Kong}
\date{\today}
\maketitle
\begin{abstract}
 The parity of a bit string of length $N$ is a global quantity that can be
 efficiently compute using a global counter in $\mbox{O} (N)$ time. But is it
 possible to find the parity using cellular automata with a set of local rule
 tables without using any global counter? Here, we report a way to solve this
 problem using a number of $r=1$ binary, uniform, parallel and deterministic
 cellular automata applied in succession for a total of $\mbox{O} (N^2)$ time.
\end{abstract}
\medskip
\begin{multicols}{2}
 Cellular automaton (CA) is a simple local interaction model used to study the
 evolution and self-organization of various physical and biological systems
 \cite{CABook}. And at the same time, CA can also be viewed as a restrictive
 model of parallel computation without common global memory. In fact, recently
 there is an increasing interest in using CA to perform certain computational
 tasks \cite{Comp}. It is therefore natural to ask if CA can be used to perform
 a task that depends on the global information of an input state.
\par
 An example of this kind is called the density classification problem (DCP). In
 this problem, we are given an one dimensional array of bit string in periodic
 boundary conditions. We are required to apply some CA rules so as to evolve
 the state to all zero if the number of zeros in the input state is greater
 than that of ones. Similarly, we have to evolve the state to all one if the
 number of zeros in the input state is less than that of ones.
\par
 After a number of fail attempts, Land and Belew proved that no single CA can
 solve the DCP without making some mistake \cite{LandBelew}. And yet later on,
 Capcarrere \emph{et al.} argued that a single CA rule can solve the DCP
 provided that we modify both the required output of the automaton and the
 boundary conditions used \cite{Capcarrere,Sipper}. Nonetheless, we do not
 completely agree with their approach for we have to scan through the states of
 the entire final bit string, in general, before knowing the answer. This
 requires either global memories or a table that scales with the system size in
 the read out; hence it somehow defeats the purpose of restricting ourselves to
 the use of CA in the first place. (In contrast, the read out in the original
 DCP can be determined by looking at any one or two bits in the final string.)
 Although DCP cannot be solved using one CA rule, Fuk\'{s} showed that this can
 be done by applying two CA rules in succession. More precisely, he found a
 solution to the DCP by applying the CA rule~184 a fixed number of times
 depending only on the string length and then followed by the CA rule~232 a
 fixed number of times again depending only on the string length \cite{Fuks}. A
 number of related problems depending on the global density of a string have
 also found to have CA solutions \cite{Others}.
\par
 Another challenge for CA is the parity problem (PP), namely, to evolve a given
 input bit string $\sigma$ using a sequence of CA rules to all $P_\sigma$s
 where $P_\sigma$ is the parity of $\sigma$. (Then the parity of the input bit
 string can be determined by looking at any one of the final bit string.) This
 problem appears to be much harder than the DCP because the output is altered
 simply by a flip in any one of the input bits. In fact, Sipper proved that no
 single $r=1$ CA rule can solve the PP with fixed boundary conditions
 \cite{Nonuniformity}.
\par
 In this letter, we show that PP can be solved by applying a number of $r=1$ CA
 rules in succession. The term CA in this letter shall mean a local synchronous
 uniform deterministic binary CA rule with parallel update in periodic boundary
 conditions. That is, the state of each bit in the next time step depends
 deterministically only on the state of its finite neighborhood. The states of
 all sites are updated in parallel and the rule table of the CA is covariant
 under translation along the bit string. Besides, we restrict ourselves to
 consider only those sequence of CAs that solves the PP exactly without making
 any misclassification.
\par
 To have a feeling of the difficulty of this problem, two remarks are in place.
 First, except for the boundary conditions, relaxing any of the above
 conditions makes the PP trivial. For example, applying the CA rule~60 first to
 the 2nd bit, then the 3rd bit and so on till the last bit in the bit string,
 the value of the last bit is the parity of our input string. Second, one set
 of local CA rules is not sufficient to solve the PP for input string of
 \emph{any} length. The reason is simple: if such a set could solve the PP, it
 would evolve an odd parity bit string $\sigma$ to all one. The uniformity
 condition implies that this set would evolve the concatenated even parity bit
 string $\sigma\sigma$ to all one as well. But this is absurd.
\par
 Thus, we need to invoke more than one CA rules to solve the PP. In fact, all
 we need is a few $r=1$ CA rules reported below in Wolfram's notations
 \cite{Wolfram}.
\begin{itemize}
\item $R_{222}$ is the Wolfram elementary CA rule~222.
 \begin{center}
  $R_{222}$:
  \begin{tabular}{cccccccc}
   000 & 001 & 010 & 011 & 100 & 101 & 110 & 111 \\ \hline
    0  &  1  &  1  &  1  &  1  &  0  &  1  &  1
  \end{tabular}
 \end{center}
 This rule replaces the two ending zeros of a string of zeros by ones, if the
 number of zeros in the string is more than two. Since we are using periodic
 boundary conditions, the (global) parity of the configuration does not change
 although locally parity does change. If we apply the rule $\lfloor N/2\rfloor$
 times, then there will be no more consecutive zeros.
\item $R_{132}$ is similar to $R_{222}$, but it replaces the ending ones by
 zeros. It also conserves parity.
 \begin{center}
  $R_{132}$:
  \begin{tabular}{cccccccc}
   000 & 001 & 010 & 011 & 100 & 101 & 110 & 111 \\ \hline
    0  &  0  &  1  &  0  &  0  &  0  &  0  &  1
  \end{tabular}
 \end{center}
\item $R_{76}$ will flip a configuration of all ones to all zeros.
 \begin{center}
  $R_{76}$:
  \begin{tabular}{cccccccc}
   000 & 001 & 010 & 011 & 100 & 101 & 110 & 111 \\ \hline
    0  &  0  &  1  &  1  &  0  &  0  &  1  &  0
  \end{tabular}
 \end{center}
\item $R_{254}$ preserves the configuration of all zeros. In fact, careful
 application of $R_{254}$ together with $R_{76}$ will put our result to the
 desired form.
 \begin{center}
  $R_{254}$:
  \begin{tabular}{cccccccc}
   000 & 001 & 010 & 011 & 100 & 101 & 110 & 111 \\ \hline
    0  &  1  &  1  &  1  &  1  &  1  &  1  &  1
  \end{tabular}
 \end{center}
\item $R_{184}$ is the so-called traffic rule. It tries to move an one to the
 right if the site at right is a zero.
 \begin{center}
  $R_{184}$:
  \begin{tabular}{cccccccc}
   000 & 001 & 010 & 011 & 100 & 101 & 110 & 111 \\ \hline
    0  &  0  &  0  &  1  &  1  &  1  &  0  &  1
  \end{tabular}
 \end{center}
\item $R_{252}$ is an auxiliary rule to change a zero right of an one to one.
 \begin{center}
  $R_{252}$:
  \begin{tabular}{cccccccc}
   000 & 001 & 010 & 011 & 100 & 101 & 110 & 111 \\ \hline
    0  &  0  &  1  &  1  &  1  &  1  &  1  &  1
  \end{tabular}
 \end{center}
\end{itemize}
\par
 Let $\sigma$ be an arbitrary input bit string of length $N$, and $P_\sigma$
 its parity. We write $R\sigma$ the resulting configuration after the rule $R$
 is applied to $\sigma$ once, for any CA rule $R$.
\par\medskip\noindent
\emph{Theorem}. Let
\begin{equation}
 \sigma_1\equiv \left({R_{132}}^{\lfloor\frac{N}{2}\rfloor}
                     {R_{222}}^{\lfloor\frac{N}{2}\rfloor}
               \right)^{\lfloor\frac{N}{2}\rfloor}\sigma~,
\end{equation}
 which means that we apply $R_{222}$ to $\sigma$ $\lfloor N/2\rfloor$ times,
 then apply $R_{132}$ $\lfloor N/2\rfloor$ times and back to $R_{222}$ and so
 on. The total number of time steps is $2\lfloor N/2\rfloor^2$.\\
(a) If $N$ is odd, then
\begin{equation}
 \sigma_1=\left\{\begin{array}{ll}
                   0^N&\mbox{if $P_\sigma=0$,}\\
                   1^N&\mbox{if $P_\sigma=1$.}
                  \end{array}\right.
\end{equation}
 where $0^N$ denotes a string of $N$ consecutive zeros, that is, the state at
 all the sites is zero; and similarly for $1^N$. Total number of time steps is
 $(N-1)^2/2$.\\
(b) If $N=2q$, where $q$ is odd, then
\begin{equation}
 \sigma_1=\left\{\begin{array}{ll}
                   0^N \mbox{\ or\ } 1^N&\mbox{if $P_\sigma=0$,}\\
                   \mbox{other than $0^N$ and $1^N$} &\mbox{if $P_\sigma=1$.}
                  \end{array}\right.
\label{twoq}
\end{equation}
 Applying the CA rules $S\equiv {R_{254}}^{\lceil N/2\rceil} R_{76}$, we can
 transform $\sigma_1$ to the form given in (a).
\begin{equation}
 S\,\sigma_1=\left\{\begin{array}{ll}
                   0^N&\mbox{if $P_\sigma=0$,}\\
                   1^N&\mbox{if $P_\sigma=1$.}
                  \end{array}\right.
\end{equation}
 The total number of time steps is $(N^2/2+N/2+1)$.\\
(c) If $N=2^mq$, where $m\ge 2$ and $q$ is odd, let
\begin{equation}
 T\equiv{R_{132}}^{\lfloor\frac{N}{2}\rfloor}
         {R_{222}}^{\lfloor\frac{N}{2}\rfloor}R_{184}R_{252}
\end{equation}
and
\begin{equation}
 \sigma_2\equiv T^{m-1}\sigma_1~,
\end{equation}
then
\begin{equation}
 \sigma_2=\left\{\begin{array}{ll}
                   0^N \mbox{\ or\ } 1^N&\mbox{if $P_\sigma=0$,}\\
                   \mbox{other than $0^N$ and $1^N$} &\mbox{if $P_\sigma=1$.}
                  \end{array}\right.
\end{equation}
 Similar to (b), $S\,\sigma_2$ is either all zeros or all ones, depending on
 whether $P_\sigma$ is even or odd. The total number of steps is
 $N^2/2+(m-1)(N+2)+N/2+1$.
\par\medskip\noindent
\emph{Remark}. If we are allowed to change the lattice size, then we do not
 need (b) and (c) of the Theorem to solve the parity classification problem.
 Suppose we are given a bit string with even length. To find out the parity of
 the number of ones in it, we just concatenate a single zero bit to the bit
 string. The resultant bit string will be in odd length and hence we are back
 to (a) of the Theorem.
\par\medskip
 The following two Lemmas are required to prove this Theorem.
\par\medskip\noindent
\emph{Lemma 1}. For any $N$, $\sigma_1$ could only be one of the following
 three forms: $0^N$, $1^N$ or $(10^{2l-1})^k$, where $l,k\ge 1$ are integers,
 $2lk=N$ and $P_\sigma = k \pmod 2$. (The notation $(10^{2l-1})^k$ means, for
 example if $l=2$ and $k=3$, the bit string $100010001000$.)
\par\noindent
\emph{Proof}. It is obvious that both $0^N$ and $1^N$ are fixed points of
 $R_{222}$, $R_{132}$. Thus, in the rest of this proof, we shall only consider
 configurations $\sigma$ with both zeros and ones.  
\par
 As we have discussed above, for any configuration $\sigma$, the string 
 ${R_{222}}^{\lfloor\frac{N}{2}\rfloor}\sigma$ has no consecutive zeros. Thus,
 its general form will look like
\begin{equation}
 {R_{222}}^{\lfloor\frac{N}{2}\rfloor}\sigma = \cdots
 101^{n_1}01^{n_2}01\cdots~,
\end{equation}
 where the numbers of ones between two zeros $n_i$ are greater than or equal
 one. (The trivial configuration $1^N$ is also possible.)
\par
 Since $R_{132}$ replaces the ending ones by zeros, if any of the $n_i$ is
 even, the corresponding string of ones will be completely annihilated after
 applying ${R_{132}}^{\lfloor\frac{N}{2}\rfloor}$. For example,
\begin{eqnarray}
 {R_{132}}^2 (\cdots 0101111010\cdots) & = & R_{132} (\cdots 0100110010\cdots)
 \nonumber \\
 & = & \cdots 0100000010\cdots~.
\end{eqnarray}
 This example also shows that after a single pass of $R\equiv
 {R_{132}}^{\lfloor\frac{N}{2}\rfloor}{R_{222}}^{\lfloor\frac{N}{2}\rfloor}$,
 we cannot conclude that the number of zeros between two ones is odd. However,
 we can conclude that if there is a string of even number of zeros between two
 ones, after a single pass of $R$, the number of zeros in that string will
 increase by at least two. Thus, $R^{\lfloor\frac{N}{2}\rfloor}\sigma$ must be
 of the form
\begin{equation}
 \cdots10^{2n'_1+1}10^{2n'_2+1}1\cdots~, \label{E:form}
\end{equation}
 where $n'_i$ are non-negative integers (or the trivial cases $0^N$ or $1^N$).
\par
 We still need to show that the $n'_i$ are all equal. This can be
 illustrated by an example, consider the case $N=8$,
\begin{eqnarray}
 &&{R_{132}}^{\lfloor\frac{N}{2}\rfloor}{R_{222}}^{\lfloor\frac{N}{2}\rfloor}
  (10100000) \nonumber \\
 &=&{R_{132}}^{\lfloor\frac{N}{2}\rfloor}
  (10111011) \nonumber \\
 &=&(00010001)~.
\end{eqnarray}
 We see that the numbers of zeros between the ones tend to equalize, from one
 and five zeros to three zeros in this example. More precisely,
 ${R_{222}}^{\left\lfloor \frac{N}{2} \right\rfloor}
 (10^{2n'_1+1}10^{2n'_2+1}1\cdots)$ equals
 $01^{n'_1+n'_2+1}01^{n'_2+n'_3+1}0\cdots$ up to a translation. So, after
 applying $R$ for sufficiently long time, the resultant string will be in the
 form $1^N$, $0^N$ or Eq.~(\ref{E:form}). Since the first two cases are fixed
 points of $R$, we only need to focus on the third case. If the resultant
 string is in the third form, then it is easy to check that the total number of
 ones does not increase after each application of $R$. Hence, eventually the
 number of ones in the string will stay constant under repeated application of
 $R$. And this happens if and only if $n'_i$ are all even or all odd. In this
 case,
\begin{eqnarray}
 & & R (10^{2n'_1+1}10^{2n'_2+1}1\cdots) \nonumber \\
 & = & 10^{1+n'_2+(n'_1+n'_3)/2}10^{1+n'_3+(n'_2+n'_4)/2}1\cdots
\end{eqnarray}
 up to a translation. Hence, the repeated application of $R$ equalizes the
 number of zeros between the ones; and a configuration in the form $\left(
 10^{2n+1} \right)^l$ is a fixed point of $R$. Finally, it is straight forward
 to check that at most $\left\lfloor \frac{N}{2} \right\rfloor$ applications of
 $R$ is enough to bring a bit string to the fixed points of the form $0^N$,
 $1^N$ or $(10^{2l-1})^k$. (One of the worst cases is the configuration
 $1(10)^{(N-1)/2}$ for $N$ odd.)
\par
 Since $R_{132}$ and $R_{222}$ conserve parity, $k \pmod 2$ is equal to the
 parity of $(10^{2l-1})^k$, which is just $P_\sigma$. This completes the proof
 of Lemma~1.
\hfill$\Box$
\par\medskip\noindent
\emph{Lemma 2}. With the notations of the Theorem, we have
\begin{eqnarray}
  T\,0^N &=& 0^N \label{T0}\\
  T\,1^N &=& 1^N \label{T1}\\
  T\,(10^{2l-1})^k &=&
    \left\{\begin{array}{ll}
            (10^{l-1})^{2k} & \mbox{if $l$ is even,}\\
            1^N             & \mbox{if $l=1$,}\\
            0^N             & \mbox{if $l\ge 3$ is odd,}
           \end{array}\right. \label{Tother}
\end{eqnarray}
 where $k\ge1$.
\par\noindent
\emph{Proof}. Eq.~(\ref{T0}) and Eq.~(\ref{T1}) are trivial. For
 Eq.~(\ref{Tother}), we consider two cases. If $l\ge2$,
\begin{eqnarray}
   T\,(10^{2l-1})^k
  &=& \left({R_{132}}^{\lfloor\frac{N}{2}\rfloor}
            {R_{222}}^{\lfloor\frac{N}{2}\rfloor}
             R_{184}R_{252}\right)\,(10^{2l-1})^k \nonumber \\
  &=& \left({R_{132}}^{\lfloor\frac{N}{2}\rfloor}
            {R_{222}}^{\lfloor\frac{N}{2}\rfloor}
             R_{184}\right)\,(110^{2l-2})^k \nonumber \\
  &=& \left({R_{132}}^{\lfloor\frac{N}{2}\rfloor}
            {R_{222}}^{\lfloor\frac{N}{2}\rfloor}\right)\,
             (1010^{2l-3})^k \nonumber \\
  &=& \left\{\begin{array}{ll}
              (10^{l-1}10^{l-1})^k & \mbox{if $l$ is even,}\\
              (0^{2l})^k           & \mbox{if $l$ is odd.}
             \end{array}\right.
\end{eqnarray}
If $l=1$,
\begin{eqnarray}
   T\,(10)^k
  &=& \left({R_{132}}^{\lfloor\frac{N}{2}\rfloor}
            {R_{222}}^{\lfloor\frac{N}{2}\rfloor}
             R_{184}R_{252}\right)\,(10)^k \nonumber \\
  &=& \left({R_{132}}^{\lfloor\frac{N}{2}\rfloor}
            {R_{222}}^{\lfloor\frac{N}{2}\rfloor}
             R_{184}\right)\,1^N \nonumber \\
  &=& 1^N~.
\end{eqnarray}
 This concludes the proof of Lemma~2.
\hfill$\Box$
\par\medskip\noindent
\emph{Proof of the Theorem}. (a)~If $N$ is odd, among the three forms provided
 by Lemma~1, $(10^{2l-1})^k$ cannot be reached. Since parity is conserved, the
 final configuration can only be $0^N$ or $1^N$ according to the initial
 parity.
\par
 (b)~If $N=2q$, parity of $0^N$ and $1^N$ are even, but the parity of
 $(10^{2l-1})^k$ must be odd because $k$ is odd. Thus, we have proved
 Eq.~(\ref{twoq}). Now, by flipping the all ones to all zeros and changing all
 other configurations to all ones by the rules $S$, we put the final
 configuration to a form similar to (a).
\par
 (c)~If $P_\sigma=1$, then $P_\sigma=1=k \pmod 2$ and $l=2^{m-1}q'$ where $q'$
 is odd. By Lemma~2, we have
\begin{eqnarray}
  && T^{m-1}(10^{2l-1})^k =  T^{m-1}(10^{2^mq'-1})^k \nonumber \\
 &=& T^{m-2}(10^{2^{m-1}q'-1})^{2k} \nonumber \\
 &=& \cdots \nonumber \\
 &=& (10^{2q'-1})^{2^{m-1}k}~.
\end{eqnarray}
 If $P_\sigma=0$, then $l=2^{m'-1}q'$ where $m'<m$ and $q'$ is odd. The above
 equation does not reach the last line and $T^{m-1}(10^{2l-1})^k$ is $0^N$ or
 $1^N$. This completes the proof of the Theorem.
\hfill$\Box$
\par\medskip
 To summarize, we show that it is impossible for any single set of CA rules to
 correctly compute the parity of a bit string. But surprisingly, we find a CA
 solution to the PP using a sequence of $r=1$ CAs. For an input bit string of
 length $N$, the worst case run time scales as $\mbox{O} (N^2)$. However, we
 have no idea if the present method has the shortest worst case run time or
 not. And since the PP is equivalent to the computation of the sum of a given
 bit string modulo two, our result implies that CA can be used to count the
 number of ones in a bit string modulo two. In other words, if we denote the
 number of ones in a bit string $\sigma$ by $\#\sigma$, then one can use a
 sequence of CAs to compute the least significant bit of $\#\sigma$ written in
 binary notation. Besides, the CA solution of the DCP for a bit string $2^n-1$
 long can then be regarded as a way to compute the most significant bit of
 $\#\sigma$. Therefore, it is instructive to investigate the possibility of
 using CAs to compute any given bit of $\#\sigma$ and hence to count the number
 of ones in a bit string provided that we have $\mbox{O} (\log N)$ copies.
 Along a similar line of thought, it is also worthwhile to look for CA
 solutions to the problem of addition over a finite ring or field.
\acknowledgements
 This work is supported in part by the Hong Kong SAR Government RGC grant
 HKU~7098/00P. H.F.C. is also supported in part by the Outstanding Young
 Researcher Award of the University of Hong Kong.

\end{multicols}

\begin{references}
\bibitem{CABook} See, for example, B. Chopard and M. Droz, {\it Cellular
 Automata Modeling Of Physical Systems} (CUP, Cambridge, 1998).
\bibitem{Comp} M. Mitchell, P.~T. Hraber and J.~P. Crutchfield, Complex\ Sys.
 {\bf 7}, 89 (1993); S.~C. Benjamin and N.~J. Johnson, \apl {\bf 70}, 2321
 (1997); M. Sipper, Computer {\bf 32}, 18 (1999).
\bibitem{LandBelew} M. Land and R.~K. Belew, \prl {\bf 74}, 5148 (1995).
\bibitem{Capcarrere} M.~S. Capcarrere, M. Sipper and M. Tomassini, \prl
 {\bf 77}, 4969 (1996).
\bibitem{Sipper} M. Sipper, M.~S. Capcarrere and E. Ronald,
 Int.\ J.\ Mod.\ Phys.\ C\ {\bf 9}, 899 (1998).
\bibitem{Fuks} H. Fuk\'{s}, \pre {\bf 55}, R2081 (1997).
\bibitem{Others} H.~F. Chau, K.~K. Yan, K.~Y. Wan and L.~W. Siu, \pre {\bf 57},
 1367 (1998); H.~F. Chau, L.~W. Siu and K.~K. Yan,
 Int.\ J.\ Mod.\ Phys.\ C\ {\bf 10}, 883 (1999).
\bibitem{Nonuniformity} M. Sipper, \pre {\bf 57}, 3589 (1998).
\bibitem{Wolfram} S. Wolfram, \rmp {\bf 55}, 601 (1983).
\end{references}
\end{document}